\documentclass{phbauth}
\usepackage{graphicx}

\begin{document}

\begin{frontmatter}

\title{Vortex state structure  of a Bose condensate in an asymmetric
trap }
\author[address1]{Anatoly A.~Svidzinsky\thanksref{thank1}},
\author[address1]{Alexander L.~Fetter}

\address[address1]{Department of Physics, Stanford University,
Stanford, CA 94305-4060}

\thanks[thank1]{Corresponding author. E-mail: asvid@leland.Stanford.EDU}

\begin{abstract}
Based on an analytic solution of the Gross-Pitaevskii equation in the
large-condensate (Thomas-Fermi) limit we determine the structure of a
 stationary vortex in a Bose-Einstein condensate in a nonaxisymmetric 
rotating trap. The  condensate velocity field has
cylindrical symmetry only near the vortex core and becomes intrinsically
anisotropic near the condensate boundaries. Rotating the anisotropic trap
induces  an additional irrotational velocity field even for a vortex-free
condensate.
\end{abstract}


\end{frontmatter}









The experimental achievement of Bose-Einstein condensation in confined
alkali-atom gases \cite{And,Brad1,Dav} has stimulated great interest in the
generation and observation of vortices in such systems \cite{JMA,DCLZ,MZW}.
Rotating a totally  anisotropic harmonic trap at an
angular frequency $\Omega $
 can, in principle,  generate vortices; they are  energetically
stable for
$\Omega >\Omega _c$ \cite{BP,D,SF,FCS}.

This work determines the condensate velocity field for a
stationary vortex  when the completely anisotropic harmonic
trap potential $V_{\rm tr}=\frac 12 M(\omega _x^2x^2+\omega
_y^2y^2+\omega _z^2z^2)$ rotates about a principal axis. We consider a
zero-temperature condensate in the Thomas-Fermi (TF) limit, when the vortex
core radius $\xi \sim d^2/R$ is small compared to the mean
oscillator length $d$ and the mean dimension $R$ of the condensate [here,
$d=(d_xd_yd_z)^{1/3}$ with $d_i =\sqrt{
\hbar /M\omega_i}$  and trap frequencies $\omega_i $ ($i =x, y, z$)].

We assume that the trap rotates with an angular velocity $\Omega $ around the
$z$ axis. At zero temperature in the rotating frame, the equilibrium
condensate wave function $\Psi$ satisfies the Gross-Pitaevskii equation:
\begin{equation}
\label{1}\left( -\frac{\hbar ^2}{2M}\nabla ^2+V_{{\rm tr}}+g|\Psi |^2-\mu
+i\hbar \Omega \partial _\phi \right) \Psi =0,
\end{equation}
where $g=4\pi \hbar ^2a/M>0$ is the effective interparticle interaction
strength, $\mu $ is the chemical potential in the rotating frame, and $\phi$
is the azimuthal angle in cylindrical polar coordinates ($r,\phi,z$).
Consider a $q$-fold vortex, oriented along the $z$ axis, with a
condensate wave function $\Psi =|\Psi |e^{iS}$.  Its phase must have the
 form
$S=q\phi +S_1$, where $S_1$ is an odd  $2\pi $-periodic function of $\phi $.

Multiplying  the imaginary part of (\ref{1})  by $|\Psi |$ yields
the  equation of current continuity in the rotating frame:
\begin{equation}
\label{5}{\rm div}\left( |\Psi |^2\nabla S\right) -\frac M\hbar \Omega \partial
_\phi |\Psi |^2=0.
\end{equation}
The second term here arises from the difference between the condensate velocity
field
${\bf v}^l=\frac
\hbar M\nabla S$ in the laboratory frame and  ${\bf v}^r=\frac \hbar
M\nabla S-{\bf
\Omega }\times {\bf r}$  in the rotating frame. In the TF limit in the
absence of a vortex, $ g|\Psi _{TF}|^2=\mu -V_{\rm tr},$ and for distances far
from the vortex core ($ r\gg |q|\xi $) we may substitute $|\Psi _{TF}|$ into
(\ref{5}) instead of $ |\Psi |$, getting
$$
\frac{2g}M|\Psi _{TF}|^2\nabla^2 S-\big[\omega _x^2+\omega _y^2+(\omega
_x^2-\omega _y^2)\cos (2\phi )]r\partial _rS
$$
$$
+(\omega _x^2-\omega _y^2)\sin (2\phi )\!\left( \partial _\phi S-\!\frac{
M\Omega }\hbar r^2\right) \!-2\omega _z^2z\partial _zS=0.
$$

Let us seek a solution of this equation in the form:
\begin{equation}
\label{7}S=S_0+q\phi -\beta \Omega r^2\sin (2\phi )
\enspace,
\end{equation}
where $\beta =(M/2\hbar) \left( \omega _x^2-\omega _y^2\right)
/\left(\omega _x^2+\omega _y^2\right)  $.
The resulting equation for $S_0$  does not contain $\Omega $. Further,
let us consider distances $x,y,z\ll R_x,R_{y,}R_z$, where $R_i$ are
the TF dimensions of the condensate ($M\omega _i^2R_i^2=2\mu $, $i=x,y,z$). At
these intermediate distances we can omit terms of higher order in the
parameter
$\left( x_i/R_i\right) ^2$. As a result, we find the following
inhomogeneous equation:
\begin{equation}
\label{10}\nabla^2 S_0+\left( \frac 1{R_x^2}-\frac 1{R_y^2}\right) q\sin
(2\phi )=0\enspace.
\end{equation}
This equation has an explicit  periodic  solution
\begin{equation}
\label{13}S_0=-\alpha qr^2\ln \left( \frac rA\right) \sin (2\phi)\enspace,
\end{equation}
where $\alpha = \frac 14\left( R_x^{-2}-R_y^{-2}\right)$ and
$A$ is a constant of integration that must be chosen to minimize the total
energy. We estimate
 $A\sim R_{\perp }$, where $R_{\perp }^2=2R_x^2R_y^2/\left(
R_x^2+R_y^2\right) $. Thus, the phase $S$ near  the  center of a rotating
trap in the presence of a vortex  is
\begin{equation}
\label{14}S\approx q\phi -\left[ \alpha q\ln \left( \frac r{R_{\perp }}\right)
+\beta \Omega \right] r^2\sin (2\phi )\enspace.
\end{equation}

Near the vortex core, the condensate wave function and condensate
velocity possess cylindrical symmetry, while far from the vortex core the
condensate velocity  adjusts to the anisotropy of the trap and becomes
 asymmetric. The condensate velocity in
the laboratory frame becomes
$$
\frac {M{\bf v}^l}{\hbar}\approx  \left\{ \frac qr-2\left[ \alpha q\ln \left(
\frac r{R_{\perp }}\right) +\beta \Omega \right] r\cos (2\phi )\right\} {\bf
e}_\phi
$$
\begin{equation}
\label{15} -2\left[ \alpha q\ln \left( \frac r{R_{\perp }}\right)
+\beta \Omega \right] r\sin (2\phi ){\bf e}_r\,.
\end{equation}
The energy barrier to vortex formation \cite{SF} means that the condensate can
be vortex free even for $\Omega > \Omega _c$, 
in which case the trap rotation induces an irrotational velocity field ${\bf v}
^l\propto \Omega\nabla (xy)$~\cite{ALF}. To logarithmic accuracy, the
additional
 phase
$S_0$ does not affect the  critical angular velocity $\Omega_c$ and metastable
angular velocity $\Omega_m = \frac 35\Omega_c$  found previously~\cite{SF}.

In conclusion, an analytic solution of Gross-Pitaevskii equation in
the rotating frame gives the  condensate velocity field for a
vortex state in an anisotropic rotating  trap.  Even for a vortex-free
condensate, the  rotation induces superfluid  motion.

This work was supported in part by the National Science Foundation, Grant
No. DMR 94-21888, and by Stanford University (A.A.S.).


\end{document}